\documentstyle[aps,pre,epsf,floats,twocolumn]{revtex}

\begin{document}

\draft

\twocolumn[\hsize\textwidth\columnwidth\hsize\csname@twocolumnfalse\endcsname

\title{Nature of crossover from classical to Ising-like critical behavior} 
\author{Erik Luijten$^{1,2,}$\cite{email} and Kurt Binder$^2$}
\address{$^1$Max-Planck-Institut f\"ur Polymerforschung, Postfach 3148,
         D-55021 Mainz, Germany}
\address{$^2$Institut f\"ur Physik, WA 331, Johannes Gutenberg-Universit\"at,
         D-55099 Mainz, Germany}

\date{April 1, 1998}

\maketitle

\begin{abstract}
  We present an accurate numerical determination of the crossover from
  classical to Ising-like critical behavior upon approach of the critical point
  in three-dimensional systems. The possibility to vary the Ginzburg number in
  our simulations allows us to cover the entire crossover region. We employ
  these results to scrutinize several semi-phenomenological crossover scaling
  functions that are widely used for the analysis of experimental results. In
  addition we present strong evidence that the exponent relations do not hold
  between effective exponents.
\end{abstract}


]
\narrowtext

It is now well established that the critical behavior of large classes of
systems---including uniaxial ferromagnets, binary alloys, simple fluids, fluid
mixtures, and polymer blends---belongs to the three-dimensional (3D) Ising
universality class.  However, this behavior is only observed asymptotically
close to the critical point, whereas {\em classical\/} critical behavior is
(sometimes) observed at temperatures farther from the critical temperature
$T_c$, before one enters the noncritical background. At intermediate
temperatures, a continuous crossover occurs from one universality class to the
other. Due to the limited extent of the asymptotic (Ising) regime, the
thermodynamic behavior in this crossover region is extremely relevant from an
experimental point of view and hence has attracted long-standing attention.
Despite this attention, several fundamental questions concerning the crossover
are still open to debate.

First, we note that the vast majority of all relevant studies is limited to the
one-phase region only. The behavior of the susceptibility (compressibility) in
this region is described by means of several approaches, which all consist of
more or less phenomenological extensions of the renormalization-group (RG)
description of the critical behavior.  In Ref.~\cite{anisimov92} three
different descriptions have been compared. From a nonlinear treatment of
$\phi^4$~theory at fixed dimensionality $d=3$, Bagnuls and
Bervillier~\cite{bagnuls84,bagnuls85} obtained functions describing the full
crossover from asymptotic critical behavior to classical critical behavior.
Belyakov and Kiselev~\cite{belyakov} carried out a first-order
$\varepsilon$-expansion, which was then phenomenologically extended to yield
the correct asymptotic behavior in the Ising regime. Finally,
Ref.~\cite{anisimov92} discusses an extension by Chen {\em et al.\/}\ of the
work of Nicoll and Bhattacharjee~\cite{nicoll81}, based on an RG matching
technique. All three approaches yield rather similar results and suggest (for
simple fluids) a smooth monotonic crossover.

Yet, from the experimental side it has not been so easy to confirm this picture
or even to make judgments concerning the quality of the various predictions.
Widespread attention attracted the findings~\cite{corti} on micellar solutions,
in which exponents were observed that did not fit into any of the known
universality classes. This was analyzed by Fisher~\cite{fisher-eff} in terms of
a crossover effect, suggesting that a nonmonotonic variation of the
susceptibility exponent in the one-phase region is an intrinsic property of the
universal crossover scaling function.  This possibility of nonmonotonicity was
then essentially confirmed in Ref.~\cite{bagnuls87a}, although again an
empirical extension of RG theory had to be invoked as well as extremely strong
higher-order corrections. All authors stressed the need for much more accurate
experimental results and explicit calculations of the crossover behavior.
Indeed, several experiments have been carried out on polymer
blends~\cite{meier93,schwahn94}, which were analyzed in terms of the crossover
solution by Belyakov and Kiselev.  Polymer systems offer the advantage that the
Ginzburg number ruling the crossover can be varied by varying the length of the
polymer chains.  Nevertheless, it proved difficult to cover the full crossover
region and results for different polymer blends with widely varying chemical
properties had to be combined. This in turn led to conjectures concerning an
unexpected pressure dependence of the Ginzburg number $G$, which needed to be
fitted to each system separately~\cite{schwahn94}. A determination of effective
exponents has not been attempted in these studies.  More recently, Anisimov and
coworkers~\cite{anisimov95} focused on the possibility of nonmonotonic
variations of effective exponents in complex fluids, whereas also for polymer
solutions a sharp, nonmonotonic crossover of the susceptibility exponent has
been reported~\cite{melnichenko}. However, it is unclear how the results on
micellar solutions fit into this picture and also the applicability of the
crossover solution by Chen {\em et al.\/}\ to systems which exhibit a
nonmonotonic crossover is subject to some debate. Nonuniversal crossover
(depending on parameters in addition to the Ginzburg number) is
suggested~\cite{anisimov95}, but this behavior may be particular to complex
fluids due to the occurrence of mesoscopic lengths in addition to the
correlation length.

In Ref.~\cite{chicross}, we presented a numerical study of the crossover to
classical critical behavior on either side of the critical temperature in {\em
two-dimensional\/} systems. This revealed a strictly monotonic crossover of the
susceptibility exponent in the disordered phase (one-phase region) and a
nonmonotonic variation in the ordered phase. Although this was an interesting
finding in itself, it also stressed the importance of a study of 3D systems,
since only these systems can be compared with the various theoretical crossover
descriptions and also an even {\em qualitative\/} difference with the
two-dimensional case could not be excluded. It is the objective of this paper
to present the results of a major numerical effort to determine the crossover
in 3D spin systems. These results allow, for the first time, a detailed and
rigorous test of the above-mentioned crossover functions.

The crossover is ruled by the parameter $t/G$, where $t \equiv (T-T_c)/T_c$ is
the reduced temperature and $G$ the so-called Ginzburg number.  Asymptotic
critical behavior occurs for $t/G \ll 1$ and classical critical behavior is
expected for $t/G \gg 1$. The additional requirement that $t$ must be small
implies that only systems with a very small~$G$ (large interaction range) allow
an observation of the full crossover. In simple fluids, e.g., the crossover is
never completed before leaving the critical region.  Numerical calculations
offer the advantage that $G$ is known precisely and can be made arbitrarily
small by increasing the range of the interactions. Thus, we could vary $t/G$
over more than eight orders of magnitude, compared to four orders of magnitude
in experiments on polymer blends~\cite{meier93}.  The large variation of $t/G$
requires the simulation of systems with very large coordination numbers. Until
now, this constituted the main bottleneck for explicit calculations. However,
the advent of a new Monte Carlo algorithm~\cite{ijmpc} for long-range
interactions has now allowed us to cover the full crossover region.

We have simulated classical spin systems, consisting of 3D simple cubic
lattices with periodic boundary conditions. The systems are described by the
Hamiltonian ${\cal H}/k_B T = - \sum_{ij} K_d({\bf r}_i - {\bf r}_j)s_i s_j$,
where $s = \pm 1$, the sum runs over all spin pairs, and the coupling depends
on the distance $|{\bf r}|$ between the spins as $K_d({\bf r})=cR_m^{-d}$ for
$|{\bf r}| \leq R_m$ and $K_d({\bf r})=0$ for $|{\bf r}|>R_m$. For any finite
$R_m$, the critical behavior of this model belongs to the Ising universality
class, but for $R_m \to \infty$ it will be classical.  The consequent singular
dependence on $R_m$ of all critical amplitudes has been derived in
Refs.~\cite{mon,medran}. In order to avoid lattice effects we formulate our
analysis in terms of an {\em effective\/} interaction range $R$~\cite{mon}.
The $R$ dependence of the Ginzburg number is given by $G = G_0 R^{-2d/(4-d)}$,
such that the crossover occurs as a function of $tR^6/G_0$. By increasing $R$
we can reach the classical regime while still keeping $t \ll 1$ in order to
stay within the critical region. On the other hand, very small values of $t$
imply a strongly diverging correlation length and extremely large system sizes
are required to avoid finite-size effects.  Thus, we have constructed all
crossover functions by combining the results for systems with different
Ginzburg numbers, such that $t$ had to be varied only within a limited range.
Systems with linear size up to $L=160$ (ca.\ 4 million spins) have been
simulated, in which each spin interacts with up to 8408 neighbors,
corresponding to an effective interaction range of 9.8 intermolecular
distances.

We first consider the experimentally most widely studied case, {\em viz.\/}\ 
$T>T_c$. We have calculated the susceptibility from the magnetization
density~$m$, $\chi' = L^d \langle m^2 \rangle / k_B T$, and the scaled
susceptibility $\tilde{\chi}=k_B T_c(R) \chi'$.  In the Ising limit, the latter
quantity diverges as $C_I^+ t^{-\gamma}$, with $\gamma = 1.237$~\cite{ic3d} and
$C_I^+ = 1.1025$~\cite{liu_fisher}, whereas mean-field theory predicts
$\tilde{\chi} = 1/t$, i.e., $\gamma_{\rm MF}=1$.  The leading range dependence
of the critical susceptibility amplitude is given by $R^{2d(1-\gamma)/(4-d)}$
\cite{mon,medran}.  Thus, if one plots the data as a function of the reduced
variable $t/G$, a data collapse should be obtained for $\tilde{\chi} G_0 /R^6$.
We will now test whether this quantity reproduces the predicted crossover
behavior and how well it is described by various theoretical expressions.
First, we consider the phenomenological generalization of first-order
$\varepsilon$-expansion results obtained in~\cite{belyakov}.  The main
significance of this function lies in its rather widespread use for the
analysis of crossover effects in polymer systems.  It is given in the following
form,
\begin{eqnarray}
 t/G &=& [1 + \kappa (\tilde{\chi}G)^{\theta/\gamma}]^{(\gamma-1)/\theta} 
         \nonumber \\
     & & \times \left\{ (\tilde{\chi}G)^{-1} + 
         [1+\kappa(\tilde{\chi}G)^{\theta/\gamma}]^{-\gamma/\theta}\right\}\;,
\label{eq:bk}
\end{eqnarray}
where $\theta=0.508~(25)$~\cite{talapov} is Wegner's correction-to-scaling
exponent and $\kappa \approx 2.333$ a universal constant.  Asymptotically close
to the critical point, $\tilde{\chi} = G^{\gamma-1}
(1+\kappa^{\gamma/\theta})^\gamma \kappa^{-\gamma/\theta} t^{-\gamma}$ and for
$t/G \gg 1$ $\tilde{\chi}$ exhibits the limiting behavior $1/t$.  As the curve
reproduces the amplitude of the mean-field asymptote, the only remaining
adjustable parameter is a multiplicative constant in~$G$.  This is precisely
the way $G$ is determined in experimental analyses.  Considering $\tilde{\chi}$
given by (\ref{eq:bk}) as the master curve, we set $G_0=[C_I^+
\kappa^{\gamma/\theta} (1+\kappa^{\gamma/\theta})^{-\gamma} ]^{1/(\gamma-1)}
\approx 0.1027$ such that the 3D Ising asymptote coincides with the curve in
the limit $t/G \ll 1$. On the other hand, we may also calculate the {\em
exact\/} Ginzburg number from the expression given in Ref.~\cite{belyakov},
which for our model reduces to $G=27/(\pi^4 R^6) \approx 0.27718/R^6$. The
former value $G_0=0.1027$ leads to a precise reproduction of the Ising
asymptote and hence has been used in the graph of the crossover curve, but the
exact value of~$G$ may yield a better description of the overall crossover
behavior.  Secondly, we consider the crossover function resulting from the
nonlinear RG treatment of Ref.~\cite{bagnuls84}.  It is presented in terms of a
phenomenological function~$\chi^*(t/g_0^2)/g_0^2$, which represents the RG
calculations with a relative error of less than $10^{-4}$~\cite{bagnuls84}. The
merit of this field-theoretic treatment is that it has a more solid foundation
than Eq.~(\ref{eq:bk}). Also $\chi^*$ reproduces the mean-field asymptote and
thus we adjust the parameter $g_0$ such that $\chi^*(t/g_0^2)/g_0^2$ coincides
with $\tilde{\chi}$ in the Ising regime.  As mentioned before, a third
crossover function for the susceptibility exists, based on an RG matching
technique.  However, we have not included this function in the present
discussion, because of its very close resemblance to the solution of
Ref.~\cite{belyakov}.

\begin{figure}
\epsfxsize 3.0in
\epsfbox{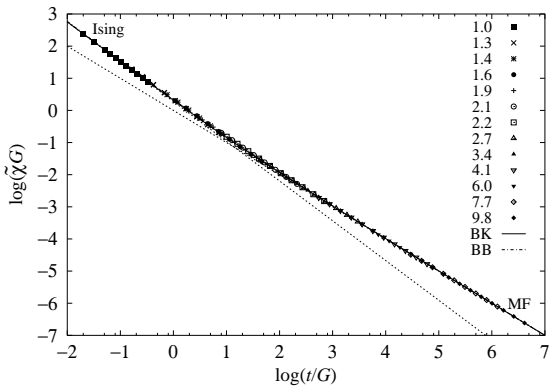}
\caption{Crossover curve for the magnetic susceptibility above~$T_c$. The
  numbers in the key refer to the effective interaction range~$R$. ``BK'' and
  ``BB'' indicate the crossover functions of Refs.~\protect\cite{belyakov}
  and~\protect\cite{bagnuls84}, respectively.}
\label{fig:chi_ht}
\end{figure}

\begin{figure}
\epsfxsize 3.0in
\epsfbox{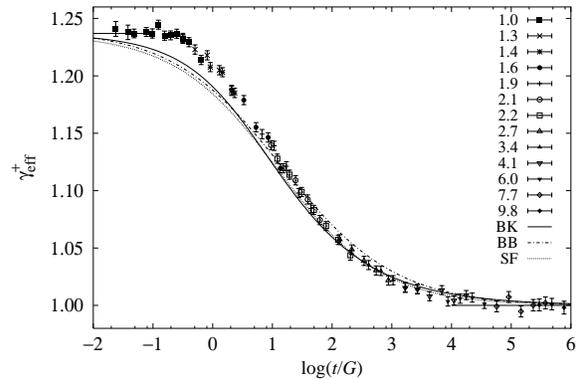}
\caption{The effective susceptibility exponent $\gamma_{\rm eff}^+$
  above~$T_c$ along with three theoretical calculations for this quantity.
  ``SF'' refers to the first-order $\varepsilon$-expansion of
  Refs.~\protect\cite{seglar,fisher-eff}.}
\label{fig:gamma_ht}
\end{figure}

Figure~\ref{fig:chi_ht} shows our results for the susceptibility in the
symmetric phase ($T>T_c$) along with the two theoretical curves.
One observes that the
simulations accurately reproduce both the Ising and the mean-field asymptote
(dotted lines). In between, the data exhibit a gradual crossover, just as
predicted by both theoretical curves. However, the vertical scale of the graph
covers ten decades, which makes it difficult to observe subtle deviations.  The
logarithmic derivative of the susceptibility, $\gamma_{\rm eff}^+ \equiv -d \ln
\tilde{\chi} / d \ln |t|$, constitutes a much more stringent test, see
Fig.~\ref{fig:gamma_ht}. The excellent collapse of the data for all our models
with widely differing interaction ranges on a master curve suggests that an
interpretation of the crossover in terms of a universal crossover function
(i.e., described by a single crossover argument $t/G$) is appropriate here. In
this graph, we have also included a third calculation by Seglar and
Fisher~\cite{seglar,fisher-eff}, which was shifted along the horizontal axis
such as to reproduce the initial deviations of $\gamma_{\rm eff}^+$ from the
mean-field value. Apart from this somewhat arbitrary shift, the curve resembles
the other predictions and is given by
\begin{equation}
  \gamma_{\rm eff}^+ = 1 + (\gamma - \gamma_{\rm MF}) E[\ln(t/G)] \;,
\label{eq:expcross}
\end{equation}
with $E(\ln y)=1/(1+y^{\varepsilon/2})$, where $\varepsilon=4-d$.  Our results
for $\gamma_{\rm eff}^+$ display a smooth, monotonic crossover from the Ising
value~$1.237$ to the classical value~$1$. While this agrees with the various
theoretical functions shown, it clearly contradicts the conjecture of
Ref.~\cite{fisher-eff}, according to which a {\em nonmonotonic\/} variation of
$\gamma_{\rm eff}^+$ in the symmetric phase might be a property of the
universal crossover scaling function.  The good description of the initial
increase of $\gamma_{\rm eff}^+$ upon approach of the critical point is an
encouraging result, since $G$ has been set to its exact value (see above) in
Fig.~\ref{fig:gamma_ht} and hence there is {\em no\/} adjustable parameter in
the solution~(\ref{eq:bk}).  However, we note that the figure also reveals a
remarkable discrepancy between the theoretical calculations and our results.
Namely, after the initial deviation from $\gamma_{\rm MF}$, the actual curve
proceeds with a considerably steeper increase than predicted by any of these
functions. Following Fisher~\cite{fisher-eff}, we define the gradient $\Gamma
\equiv -\partial \gamma_{\rm eff}^+ / \partial \log t$. Indeed, $\Gamma$
reaches values as high as 0.84 per decade, in contrast to the theoretical
functions for which the maximum of $\Gamma$ lies in the range 0.64--0.74. In
practice, this implies that the full crossover is completed {\em in one to two
decades less\/} than predicted. A final interesting aspect of
Fig.~\ref{fig:gamma_ht} is the fact that even for $R=1$ no significant
overshoot of the effective exponent above the Ising value can be observed. This
is remarkable, as it is expected~\cite{liu90} that the 3D nearest-neighbor
Ising model exhibits a {\em negative\/} leading Wegner correction. As the form
of the curve makes a further increase of $\gamma_{\rm eff}^+$ for smaller
values of~$t$ rather unlikely, we conclude that the actual effect must be very
small.

\begin{figure}
\epsfxsize 3.0in
\epsfbox{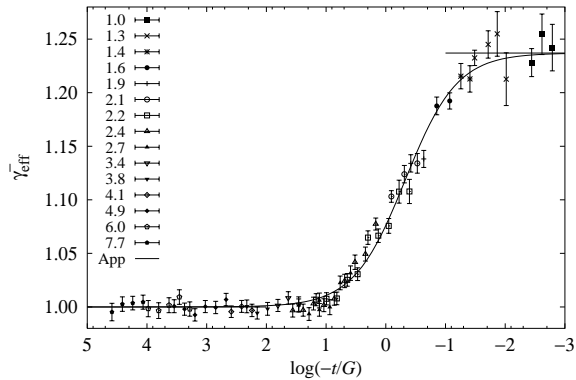}
\caption{The effective susceptibility exponent $\gamma_{\rm eff}^-$
  below~$T_c$. ``App'' denotes the approximation discussed in the text.}
\label{fig:gamma_lt}
\end{figure}

Now we proceed to the temperature region below~$T_c$. We have calculated the
susceptibility from the fluctuation relation $\chi=L^d(\langle m^2 \rangle -
\langle |m| \rangle^2)/k_B T$.
We find that the the simulation data
faithfully reproduce both the Ising and the mean-field asymptote and that the
crossover curve never deviates far from these asymptotes.  The overall graph is
very similar to that for $T>T_c$.  To our knowledge it constitutes one of the
first determinations of the full crossover function in the ordered phase.
Indeed, experimentally it is very difficult to measure the coexistence curve
and hence the crossover function.  Also from the theoretical side very few
results exist.  Bagnuls and coworkers~\cite{bagnuls87b} have carried out a
calculation using massive field theory. Just as for $T>T_c$ (see above) they
present their results in the form of an approximative continuous function.
However, this function is only valid for relatively small values of $t/G$ and
does not cover the entire crossover region. Indeed, for $t/G$ large it
approaches an asymptote with slope~$0.97$ instead of the classical value~$1$.
Thus, it is not possible to accurately describe our results with this function.
The exponent $\gamma_{\rm eff}^-$ following from our data is shown in
Fig.~\ref{fig:gamma_lt} and displays several noteworthy features.  First, also
in the ordered phase no nonmonotonicity can be observed within the statistical
accuracy, in contrast with the two-dimensional case.  Secondly, the increase of
the effective exponent upon approach of the critical point is even faster than
for $T>T_c$, with $\Gamma$ as high as~$0.11$. While $\gamma_{\rm eff}^+$
gradually starts to deviate from~$\gamma_{\rm MF}$ when $t$ is decreased, the
increase of $\gamma_{\rm eff}^-$ is rather abrupt. We view this effect as the
precursor of the ``underswing'' observed for $d=2$~\cite{chicross}. In the
absence of further calculations of $\gamma_{\rm eff}^-$, we have attempted to
describe the data by a phenomenological generalization of the exponent
crossover function~(\ref{eq:expcross}). We found that the expression $E(\ln
y)=1/(1+y)$ captures the actual behavior reasonably well (see
Fig.~\ref{fig:gamma_lt}), although the parameter~$y$ in the denominator might
carry an even larger exponent.

\begin{figure}
\epsfxsize 3.0in
\epsfbox{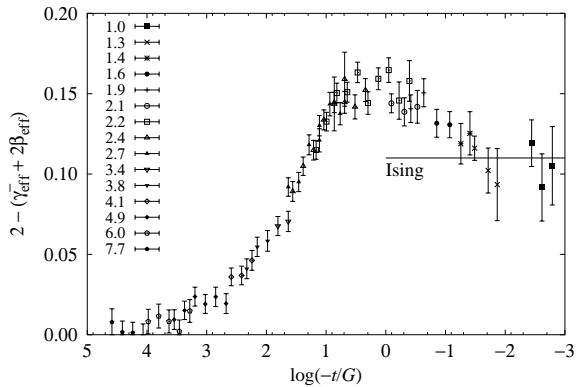}
\caption{The quantity $2-(\gamma_{\rm eff}^- + 2\beta_{\rm eff})$ as a function
  of~$t/G$.}
\label{fig:alpha}
\end{figure}

Finally, we have also considered the crossover behavior of the order parameter
$\langle |m| \rangle$ for $T<T_c$. The corresponding exponent~$\beta_{\rm eff}$
turns out to vary monotonically from its Ising value~$0.3267$ to the classical
value~$0.5$. In Fig.~\ref{fig:alpha} we display an interesting consequence of
this behavior. Namely, we have plotted the quantity $2-(\gamma_{\rm eff}^- +
2\beta_{\rm eff})$, which should be equal to the effective specific-heat
exponent $\alpha_{\rm eff}^-$ {\em if the standard scaling relations hold
between effective exponents}. However, this quantity varies between the
classical and the Ising value in a strongly nonmonotonic way, which is very
unlikely in view of the smooth behavior of $\gamma_{\rm eff}^-$ and~$\beta_{\rm
eff}$.  Thus, we consider this as strong evidence for the violation of scaling
relations between effective critical exponents. Interestingly, such a breakdown
has been inferred from $\varepsilon$-expansions over 18 years
ago~\cite{chang80}, but could experimentally not be confirmed. For the first
time it has now been explicitly demonstrated.

\acknowledgments
Stimulating discussions with Henk Bl\"ote are gratefully acknowledged.  We
thank the HLRZ J\"ulich for access to a Cray-T3E on which the computations have
been performed.


\begin{references}
\bibitem[*]{email} Electronic address: luijten@lomond.physik.uni-mainz.de
\bibitem{anisimov92} M.~A. Anisimov, S.~B. Kiselev, J.~V. Sengers, and
  S. Tang, Physica A {\bf 188}, 487 (1992).
\bibitem{bagnuls84} C. Bagnuls and C. Bervillier, J. Phys. Lett. (Paris) {\bf
  45}, L-95 (1984).
\bibitem{bagnuls85} C. Bagnuls and C. Bervillier, Phys.\ Rev.\ B {\bf 32},
  7209 (1985).
\bibitem{belyakov} M.~Y. Belyakov and S.~B. Kiselev, Physica A {\bf 190}, 75
  (1992).
\bibitem{nicoll81} J.~F. Nicoll and J.~K. Bhattacharjee, Phys.\ Rev.\ B {\bf
  23}, 389 (1981).
\bibitem{corti} M. Corti and V. Degiorgio, Phys.\ Rev.\ Lett.\ {\bf 55}, 2005
  (1985).
\bibitem{fisher-eff} M.~E. Fisher, Phys.\ Rev.\ Lett.\ {\bf 57}, 1911 (1986).
\bibitem{bagnuls87a} C. Bagnuls and C. Bervillier, Phys.\ Rev.\ Lett.\ {\bf
  58}, 435 (1987).
\bibitem{meier93} G. Meier, D. Schwahn, K. Mortensen, and S. Janssen,
  Europhys.\ Lett.\ {\bf 22}, 577 (1993).
\bibitem{schwahn94} D. Schwahn, G. Meier, K. Mortensen, and S. Janssen,
  J. Phys.\ (France) II {\bf 4}, 837 (1994).
\bibitem{anisimov95} M.~A. Anisimov, A.~A. Povodyrev, V.~D. Kulikov, and
  J.~V. Sengers, Phys.\ Rev.\ Lett.\ {\bf 75}, 3146 (1995).
\bibitem{melnichenko} Y.~B. Melnichenko {\em et al.},
  Phys.\ Rev.\ Lett.\ {\bf 79}, 5266 (1997).
\bibitem{chicross} E. Luijten, H.~W.~J. Bl\"ote, and K. Binder, Phys.\ Rev.\
  Lett.\ {\bf 79}, 561 (1997); Phys.\ Rev.\ E {\bf 56}, 6540 (1997).
\bibitem{ijmpc} E. Luijten and H.~W.~J. Bl\"ote, Int.\ J. Mod.\ Phys.\ C {\bf
  6}, 359 (1995).
\bibitem{mon} K.~K. Mon and K. Binder, Phys.\ Rev.\ E {\bf 48}, 2498 (1993).
\bibitem{medran} E. Luijten, H.~W.~J. Bl\"ote, and K. Binder, Phys.\ Rev.\ E
  {\bf 54}, 4626 (1996).
\bibitem{ic3d} H.~W.~J. Bl\"ote, E. Luijten, and J.~R. Heringa, J. Phys.\ A
  {\bf 28}, 6289 (1995).
\bibitem{liu_fisher} A.~J. Liu and M.~E. Fisher, Physica A {\bf 156}, 35
  (1989).
\bibitem{talapov} A.~L. Talapov and H.~W.~J. Bl\"ote, J. Phys.\ A {\bf 29},
  5727 (1996).
\bibitem{seglar} P. Seglar and M.~E. Fisher, J. Phys.\ C {\bf 13}, 6613 (1980).
\bibitem{liu90} A.~J. Liu and M.~E. Fisher, J. Stat.\ Phys.\ {\bf 58}, 431
  (1990).
\bibitem{bagnuls87b} C. Bagnuls, C. Bervillier, D.~I. Meiron, and B.~G. Nickel,
  Phys.\ Rev.\ B {\bf 35}, 3585 (1987).
\bibitem{chang80} M.-C. Chang and A. Houghton, Phys.\ Rev.\ Lett.\ {\bf 44},
  785 (1980).
\end{references}
\end{document}